\documentclass[letter]{aa}  

\usepackage{graphicx}
\usepackage{amsmath}
\usepackage[switch, pagewise]{lineno}

\usepackage{txfonts}
\newcommand\Lsun{L$_{\odot}$ }
\newcommand\Msun{M$_{\odot}$ }
\newcommand\Vsys{$V_{\mathrm{lsr}}$ }
\newcommand\kms{km s$^{-1}$ }
\newcommand\htco{H$_2$CO }
\begin{document} 

%\linenumbers

   \title{Molecular outflow launched beyond the disk edge}

  % \subtitle{}

\author{F.~O. Alves
\inst{1}\fnmsep\thanks{Corresponding author: Felipe O. Alves (falves@mpe.mpg.de)}
\and
J.~M. Girart \inst{2}
\and
P. Caselli\inst{1}
\and
G.~A.~P.  Franco\inst{3}
\and
B. Zhao\inst{1}
\and
W.~H.~T. Vlemmings\inst{4}
\and
M.~G. Evans\inst{5}
\and
L. Ricci\inst{6}
}

\institute{Max-Planck-Institut f\"ur extraterrestrische Physik,
Giessenbachstr. 1, D-85748 Garching, Germany
\and
Institut de Ci\`encies de l'Espai (CSIC-IEEC), Campus UAB, Carrer de 
Can Magrans S/N,
08193 Cerdanyola del Vall\`es, Catalonia, Spain
\and
Departamento de F\'isica--ICEx--UFMG, Caixa Postal 702,
30.123-970 Belo Horizonte, Brazil
\and
Department of Earth and Space Sciences, Chalmers University of Technology,
Onsala Space Observatory,
439 92 Onsala, Sweden
\and
School of Physics \& Astronomy, University of Leeds,
LS2 9LN Leeds, UK
\and
Department of Physics and Astronomy, Rice University, 
Main Street, 77005 Houston, USA
}

\date{Received May 2, 2017; accepted \today}

  \abstract
{One of the long-standing problems of star formation is the excess of angular momentum of the parent molecular cloud. 
In the classical picture, a fraction of angular momentum of the circumstellar material is removed by the magneto-centrifugally driven disk wind that is launched from a wide region throughout the disk. In this work, we investigate the 
kinematics in the envelope-disk transition zone of the Class I object BHB07-11, in the B59 core. For this purpose, we 
used the Atacama Large Millimeter/submillimeter Array in extended configuration to observe the thermal dust continuum 
emission ($\lambda_0 \sim$ 1.3 mm) and molecular lines (CO, C$^{18}$O and H$_2$CO), which are suitable tracers of 
disk, envelope, and outflow dynamics at a spatial resolution of $\sim 30$ AU. We report a bipolar outflow that was launched at 
symmetric positions with respect to the disk ($\sim$80~AU in radius), but was concentrated at a distance of 90--130~AU from 
the disk center. The two outflow lobes had a conical shape and the gas inside was accelerating. The large offset of the 
launching position coincided with the landing site of the infall materials from the extended spiral structure (seen in dust) 
onto the disk. This indicates that bipolar outflows are efficiently launched within a narrow region outside the disk edge.
We also identify a sharp transition in the gas kinematics across the tip of the spiral structure, which pinpoints the location 
of the so-called centrifugal barrier.} 


   \maketitle

\section{Introduction} 
\label{sec:intro}

The star formation process starts when the self-gravity of a dense molecular core dominates all the different 
pressure terms. The initial angular momentum in the core causes the formation of protoplanetary 
disks \citep{Boden95,Arm11}. However, a significant fraction of the initial angular momentum must be removed to allow 
the formation of stars, mainly through two outflow-launching mechanisms: (1) the X-wind, also called stellar wind, which is
 launched at a few 
stellar radii from the star \citep{Frank00}, and (2) the disk wind, which is  launched from a wide region throughout the disk
\citep{Konigl00}. The former is likely to be responsible for the collimated high-velocity jets \citep{Zinnecker98}, and the 
latter for the wide-angle outflows \citep{Bjerkeli16}. However, the exact launching position of the disk wind is not well-
constrained. \citet{Bjerkeli16} showed that the outflow-launching locations are within the inner $\sim$25~AU radius of the 
60--100~AU disk. 

The star-forming cloud B59 is located at a distance of 145 pc \citep{Alves07} and comprises the densest parts of the 
Pipe nebula, a molecular cloud that has been the subject of numerous investigations \citep[e.g.,][]{Lombardi06, 
JAlves07, Roman07, Alves08, Franco10, Frau10, Frau12}. B59 harbors a protocluster of low-mass young stellar objects (YSOs) at distinct 
evolutionary stages \citep{Onishi99,Brooke07,Forbrich09,Covey10}. Of these, BHB07-11 is the youngest member, a 
Class I object embedded in a dusty envelope with thermal emission peaking at mid-infrared bands (70 $\mu$m), and its 
luminosity is 2.7 \Lsun \citep{Brooke07,Riaz09,Forbrich09}. The large-scale kinematics ($\sim$ 4000 AU) are dominated 
by a large outflow of molecular gas perpendicular to the rotation plane of the envelope \citep{Duarte12,Hara13}.

\begin{figure}[t]
%\sidecaption
\includegraphics[width=\columnwidth,trim={2.5cm 0 3cm 0}]{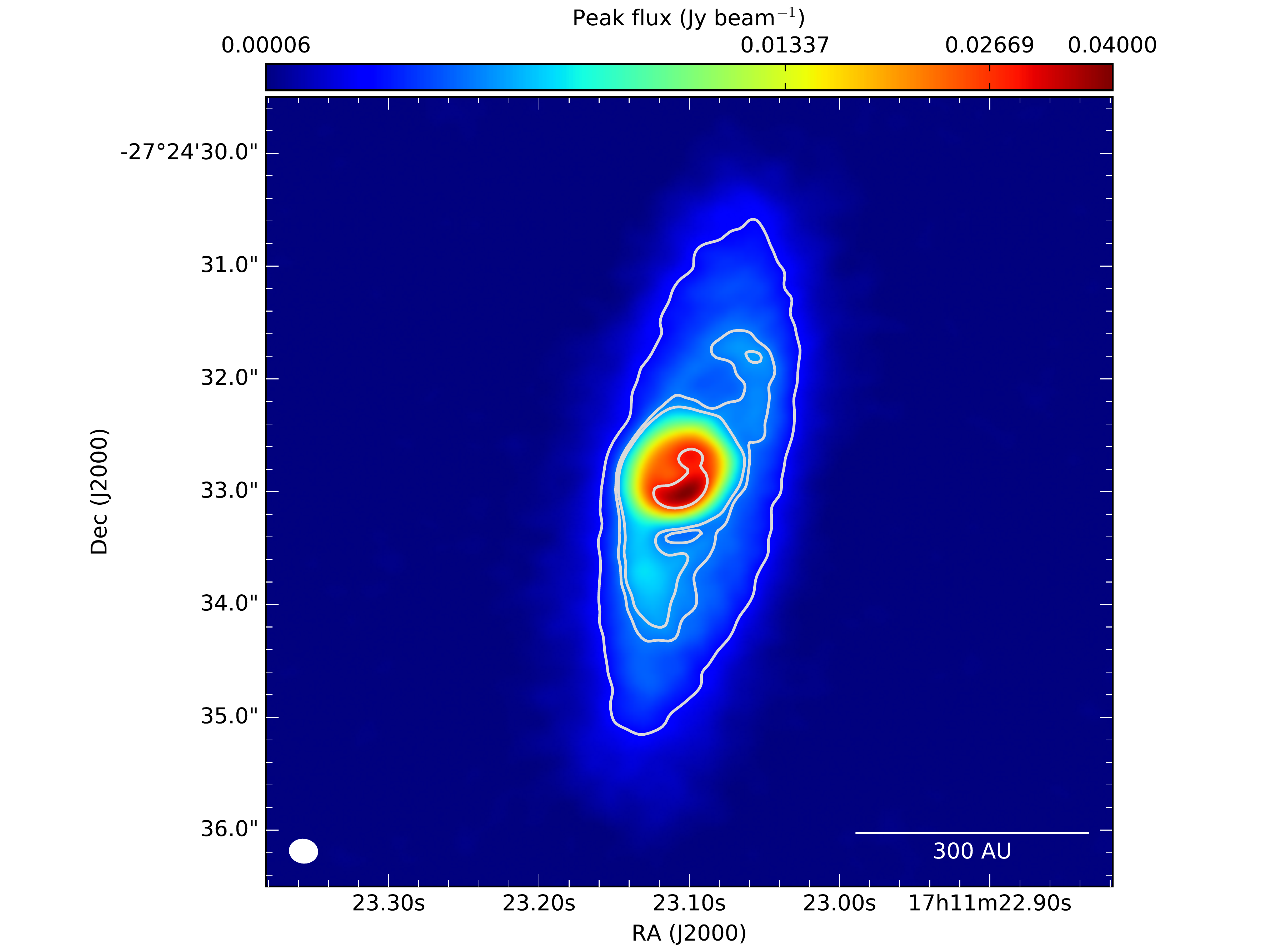}
\caption{Dust continuum emission with contours of 30, 60, 70, and 450 times the noise level 
$\sigma$ ($1\sigma = 62~\mu$Jy beam$^{-1}$) The intensity contrast between the envelope (and spirals) and the
disk is clear. 
\label{fig1}}
\end{figure}

\section{Observations}
\label{sec:obs}

The observations were carried out with the Atacama Large Millimeter/submillimeter Array (ALMA) using 44 antennas in 
an extended configuration (baselines between 0.02 and 1.6 km), at an atmospheric spectral window of 230 GHz ($
\lambda \sim 1.3$ mm). The absolute flux calibration was obtained using Titan. The bandpass calibration of the receiver 
response was performed by observing the quasar J1733$-$1304, while the relative phase and amplitude gain solutions 
were obtained from the quasar J1700$-$2610. Owing to the high variability of the calibrators, the uncertainty on the 
absolute flux density scale is about 7\%. 

The observing setup contained four spectral windows, one of which
was dedicated to continuum observations (in 
time-domain mode, TDM), and the other three dedicated to molecular line observations (in frequency-domain mode, 
FDM). The total bandwidth in continuum mode, taking into account the TDM window plus the line-free channels of the 
FDM windows, was about 2.3 GHz, leading to an {\it rms} of $62~\mu$Jy beam$^{-1}$. The molecular lines were chosen 
in order to probe gas at distinct states of temperature and density. We observed carbon monoxide (CO J = 2--1) because of 
its high abundance with respect to other molecules, C$^{18}$O (J = 2--1) and H$_2$CO (J = 3$_{0,3}$--2$_{0,2}$), 
which are tracers of dense gas (n $\sim$ 10$^{4-6}$ H$_2$ particles per cm$^3$). The spectral resolution is 0.32 km 
s$^{-1}$ for the CO data and 0.17 km s$^{-1}$ for the H$_2$CO data, with an {\it rms} noise for the line-free channels of 
0.4 and 1.5  mJy beam$^{-1}$, respectively.

The data were calibrated and imaged using CASA\footnote{The Common Astronomy Software  Applications tool by the 
ALMA staff as part of its Quality Assurance (QA) process.}. The final continuum maps were self-calibrated and improved 
by about 40\% in noise level. Our observations provide a detailed view of the disk-envelope structure and inner kinematics 
for BHB07-11. The processed maps have a spatial resolution of $\sim$ 32 AU.
   
\section{Results}
\label{sec:res}

\subsection{Dust continuum emission}
\label{subsec:dust}
The dust continuum emission reveals two distinct components exhibiting significant irregularities (Fig. \ref{fig1}). First, 
there is a bright and sharp compact disk with a radius of $\sim 80$ AU. This value is within the typical range of radii  
found in YSO disks \citep[e.g.,][]{Cotten16}. The disk is slightly elongated along a position angle (PA) of 138$\degr$ 
(east of north), and it encompasses a strong azimuthal asymmetry: the brightest emission is not at the disk center, but 
is located to the southwest of the center, forming an arc-like shape with a physical size of 50 AU.  Second, the disk is  
surrounded by a tenuous and smooth envelope\footnote{We refer to this term as envelope in a more restrictive way than usual, since this is its inner flattened component.} with a radius of 435 AU with a PA of 167$\degr$. The 
envelope contains spiral-like patterns along the major axis. Adopting the dust temperature of $\sim 31$ K measured by 
\citet{Hara13}, we estimate a disk plus envelope mass of $0.17 \pm 0.06$ \Msun \citep[assuming a dust opacity $\kappa$ of 
$0.88 \pm 0.01$ cm$^2$ g$^{-1}$ for a population dominated by grains with thin ice mantles,][]{Ossenkopf94}. The 
minor-to-major axis ratio indicates an inclination of $70\degr$ with respect to the line of sight. 

\begin{figure*}[t!]
\sidecaption
\includegraphics[width=\textwidth]{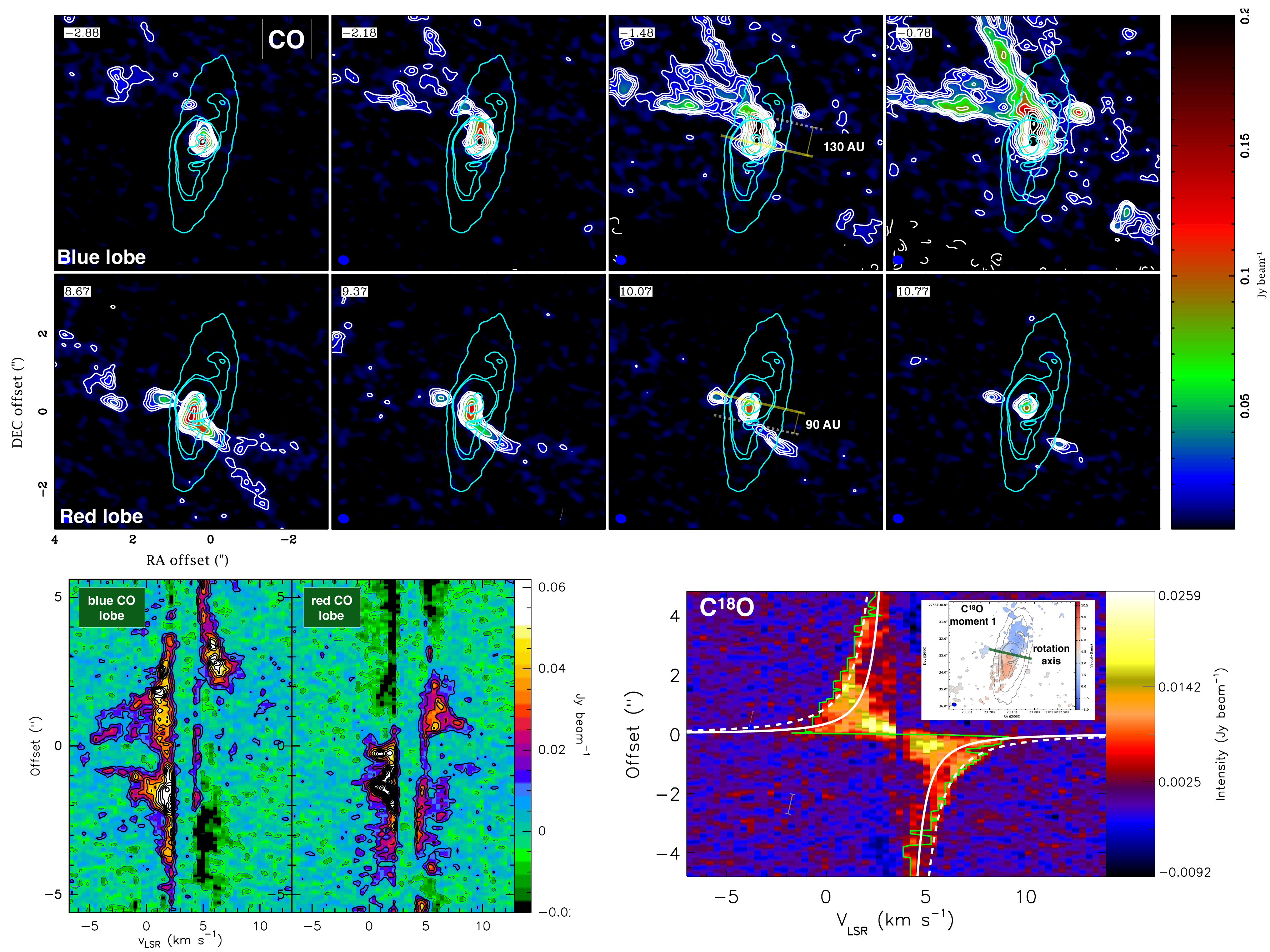}
\caption{{\it Upper panel}: CO (2--1) channel map with color-filled contours at 4, 6, 8, 10... 90$\sigma$ (1$\sigma$ = 4 
mJy beam$^{-1}$) overplotted on the dust emission (cyan contours with the same intensity levels as Fig. \ref{fig1}). The velocity 
channels are labeled in the upper left corner of each panel. The channel spacing is binned to 0.7 \kms$\!\!$. The yellow 
arrow line in channels -1.48 \kms and 10.07 \kms indicates the rotation axis of the disk. {\it Lower left panel}: PV plots 
taken from cuts transversal to the outflow axis in each lobe. The cavity pattern is indicated by the large opening angle.
{\it Lower right panel}: C$^{18}$O PV diagram from a cut along the source long axis. The dashed line shows the best 
Keplerian fit to the 3$\sigma$ envelope (green line), following the method proposed by \citet{SSW16}. The fit 
corresponds to a dynamical mass of $1.70\pm0.04$ \Msun$\!\!$, while the white line shows the fit of \citet{Hara13} for a 
mass of 0.73 \Msun$\!\!$. The inset shows the C$^{18}$O velocity field for a pixel intensity brighter than 50\% of the peak 
flux (25 mJy beam$^{-1}$). The contours refer to the continuum emission, and the green arrow shows the rotation axis. 
The synthesized beam is $0\farcs20 \times 0\farcs24$ for the
continuum, $0\farcs20 \times 0\farcs25$ for the CO data, and 
$0\farcs21 \times 0\farcs27$  for the C$^{18}$O data.
\label{fig2}}
\end{figure*}

\subsection{Molecular emission}
\label{subsec:mol}

The CO (2--1) emission at velocities in the $\pm 4$ to $\pm7$ \kms range with respect to the systemic velocity of the 
source (\Vsys $\sim 3.6$ \kms$\!\!$) shows an elongated component that arises perpendicular to the envelope major 
axis and lies onto the plane-of-sky because of the large source inclination. This molecular component originates at a 
distance of $\sim$ 90$-$130 AU from the disk center and the rotation axis, near the region of dust minimum outside the 
border of the disk (CO channel map in Fig. \ref{fig2}), where the dust spiral structures are connected to the disk. In the 
blueshifted side of the envelope, this component is extended along the northeast, whereas in the redshifted side 
it is extended in the opposite direction, along the southwest. At 90$-$130 AU, the escape velocity of the system 
is 4.8 \kms to 5.8 \kms with respect to the systemic velocity of the source (for our estimated stellar mass of
$1.70\pm0.04$ \Msun$\!\!$). This extended gas is therefore gravitationally unbound. The CO emission shows two 
features that are common in molecular outflows: it has an approximately symmetrical limb brightened conical 
cavity (lower left panel of Fig. \ref{fig2}), and there is evidence that the gas accelerates outwards. In 
both lobes, the cone vertex is offset from the rotation axis, showing that the outflow is launched far from the disk center. 
We therefore detect a molecular outflow that is launched (carrying out mass and angular momentum) outside the 
disk, far from the putative YSO. This type of outflow is different from the outflows observed so far, which are centered 
around the YSO, and it is not predicted in the classical picture of the local disk-outflow dynamics
\citep{Bland82,Shu94,Pudritz07}. The highest CO velocities ($V_{\mathrm{blue}} < -3.0$ \kms and $V_{\mathrm{red}} > 
10$ \kms$\!\!$) are confined within 70 AU of the inner disk. They suggest that BHB07-11 is a strongly perturbed disk (see 
Appedix \ref{app:high} for a brief discussion).

The C$^{18}$O (2--1) emission shows a clear velocity gradient that is fairly well aligned with the source long axis. This 
implies a rotation axis with a PA of 77$\degr$ in the plane of the sky. The position-velocity (PV) diagram pattern 
is consistent with Keplerian rotation (lower right panel of Fig. \ref{fig2}). 

The \htco (3$_{0,3}$--2$_{0,2}$) emission is confined to the dust envelope and also presents a velocity gradient with 
respect to the source center. At larger radii (120--435 AU), the spectrum profile of \htco has a pronounced asymmetry 
with redshifted absorption (upper panel in Fig. \ref{fig3}) that
is typical of infall gas motions in contracting clouds \citep[e.g.,][]
{Leung77,Myers96,Pineda12,Evans15}.

\section{Disk-envelope transition zone}
\label{sec:cb}

The transition zone between disk and envelope is crucial for understanding the disk evolution, especially for how the disk is 
assembled. The large-scale spiral structures are unlikely to originate in gravitational instabilities because the spirals 
are (1) extended structures outside the disk, and (2) not massive enough to be self-gravitating (see Appendix 
\ref{app:model}). Instead, these spirals may be infall streams spiralling inward onto the disk because most of the infall 
motion is converted into the rotation motion. In fact, we show now that the gas kinematics at the outer spiral arm shows a 
sharp transition that matches the feature of a centrifugal barrier
well.

\begin{figure*}[t]
\sidecaption
\includegraphics[width=12cm]{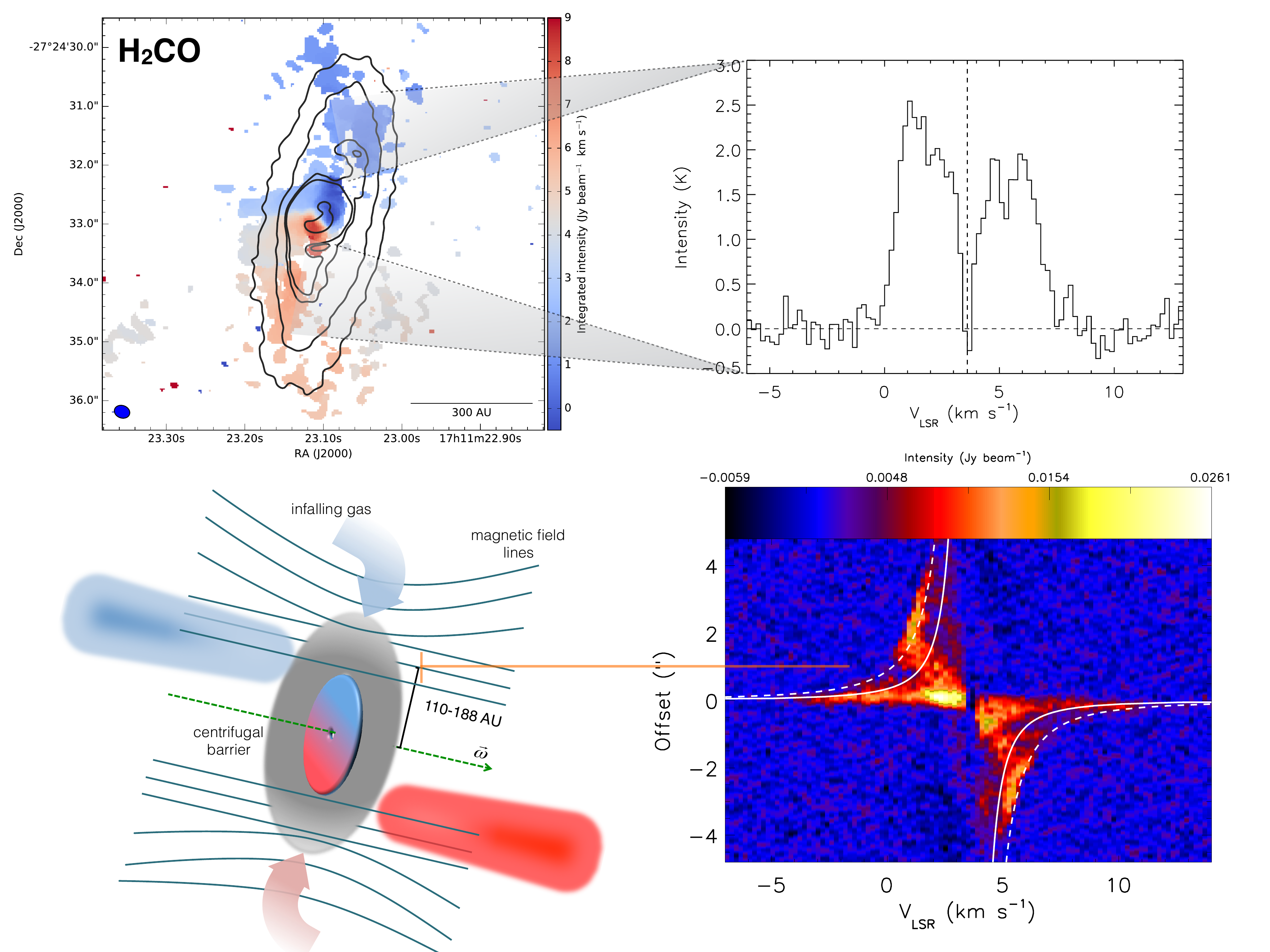}
\caption{{\it Upper panel}: Velocity field of the \htco emission with respect to the continuum emission (contours). The 
inset shows an asymmetric line profile taken from a radius between 120 and 435 AU where the blueshifted emission is 
brighter than the redshifted one, which is typical of infall motions. The thick vertical dashed line shows the ambient 
velocity according to previous observations \citep[\Vsys $\sim 3.6$ \kms$\!\!$,][]{Onishi99,Hara13}. The horizontal 
dashed line shows the continuum-removed zero-level baseline. {\it Lower panels}: Scheme showing outflows powered at 
the disk edge that are due to the magnetic field enhancement. The right panel shows the PV diagram of the \htco emission with 
the same Keplerian curves as in Fig. \ref{fig2}. The discontinuity in the velocity distribution pattern around $\sim 
1\farcs0$ indicates the change in kinematics probably caused by the centrifugal barrier. The discontinuity is preceded by a velocity peak (better seen in the northern lobe as a ``knee'' in the PV plot) and followed by a sharp increase in radial velocity in the inner 0.6$^{\prime\prime}$ as a result of the disk rotation.
\label{fig3}}
\end{figure*}

The \htco PV cut along the major axis shows a clear discontinuity between the envelope and the inner disk kinematics 
(lower right panel in Fig. \ref{fig3}). The molecular emission exceeds the Keplerian curves in velocity and reaches a 
maximum at 225~AU ($1\farcs55$) at the northern side and 121~AU ($0\farcs83$) at the southern side. It is then 
significantly decelerated between a radius of $\sim$ 110$-$188 AU ($0\farcs76$ - $1\farcs30$) at the northern side and $
\sim$ 87$-$107~AU ($0\farcs6$$-$$0\farcs74$) at the southern side. This velocity feature matches the 
centrifugal barrier in existing observational \citep[e.g.,][]{Sakai14,Oya16} and (magneto-)hydrodynamics models well
\citep{Zhao16}. The velocity peak is produced when the infalling gas piles up at their centrifugal radii. The flow dynamics 
across this centrifugal barrier is mainly dominated by rotation, while the infall motion decreases drastically. Compared 
with the dust continuum map in Fig.~\ref{fig1}, the location of the centrifugal barrier is at the tip of the spiral structures, 
with the northern lobe slightly farther out than the southern lobe.

The gas appears to flow gradually along the spiral arms onto the disk, where infall and rotation velocities are reduced, as 
indicated by the cavity feature in the PV diagram (lower right panel in Fig. 3). However, in the inner $\pm 0\farcs6$ 
region, which represents the disk, the kinematics is mainly dominated by the central protostar, and the gas resumes its 
Keplerian speed. 

It is clear that on both sides of the disk, the launching base of the main outflow coincides with the intersection of the disk 
and spiral, where the infall flow lands on the disk (Fig.~\ref{fig2}). This location is the most probable site satisfying the 
conditions of magneto-centrifgual mechanism \citep{Bland82}, because the infall drags the magnetic field lines inward 
and creates large pinching angles for the field lines \citep{Zhao16}. In comparison, field lines inside the Keplerian disk 
are much less pinched because of the rotation support and the lack of fast infall motions. Therefore, more sophisticated 
non-ideal magnetohydrodynamics (MHD) mechanisms are required for launching wind on the disk itself, which is likely to be weak and unsteady 
\citep[e.g.,][]{Bai17,Zhu17}.

Our work thus shows that the most prominent magneto-centrifugal outflows should be ejected near the landing 
site of the gas infall onto the disk, while less notable outflow components are launched in the inner radius of the disk 
(CO component aligned with the rotation axis in the redshifted channels of Fig.~\ref{fig2}). In this sense, the outflow 
observed by \citet{Bjerkeli16} is likely to belong to the weaker disk wind component; however, if the landing site of the 
gas infall in their case is not on the disk edge along the equator, but rather on the upper disk surface, it may still fit
the physical pictures we presented here. It is worth pointing out that in both observations, the outflows are not symmetric, 
as in the classical disk wind picture. The landing site of the infall streams onto the disk is indeed typically asymmetric; 
hence, symmetric bipolar outflows (not jets) should be rather rare in this framework.

\section{Conclusions}
\label{sec:conc}

This paper shows that the high-sensitivity image fidelity and angular resolution of ALMA reveals new phenomena 
associated with the formation of stars, and it shows that the dynamics of the disk-outflow systems around YSOs is 
far more complex than initially thought. In particular, outflows can be launched at a distance of $\sim$ 90$-$130 AU from 
the rotational axis, where gas infall along the extend spiral structures lands on the disk. This scenario is consistent with 
the standard magneto-centrifugal mechanism that requires a large pinching angle of the magnetic field lines, which is most 
likely to be achieved at these infall landing sites. Furthermore, the high-resolution data allow us to see the detailed 
morphologies in the transition zone of the disk and envelope, and to identify the location of the centrifugal barrier as 
on the outer spiral arm.

\begin{acknowledgements}
This paper makes use of the following ALMA data: DS/JAO.ALMA\#2013.1.00291.S. ALMA is a partnership of ESO
(representing its member states), NSF (USA) and NINS (Japan), together with NRC (Canada), NSC and ASIAA
(Taiwan), and KASI (Republic of Korea), in cooperation with the Republic of Chile. The Joint ALMA Observatory is
operated by ESO, AUI/NRAO and NAOJ. G.A.P.F. acknowledges the partial support from CNPq and FAPEMIG (Brazil). 
J.M.G. acknowledges support from MICINN AYA2014-57369-C3-P and the MECD PRX15/00435 grants (Spain). P.C.,
B.Z. and M.E. acknowledge support of the European Research Council (ERC; project PALs 320620). WV acknowledges 
support from the ERC through consolidator grant 614264. The data reported in this paper are archived in the ALMA 
Science Archive.
\end{acknowledgements}

\bibliographystyle{aa}
\bibliography{refalves}

\begin{appendix}

\section{High CO velocity components}
\label{app:high}

\begin{figure}[h!]
\centering
\includegraphics[width=\columnwidth]{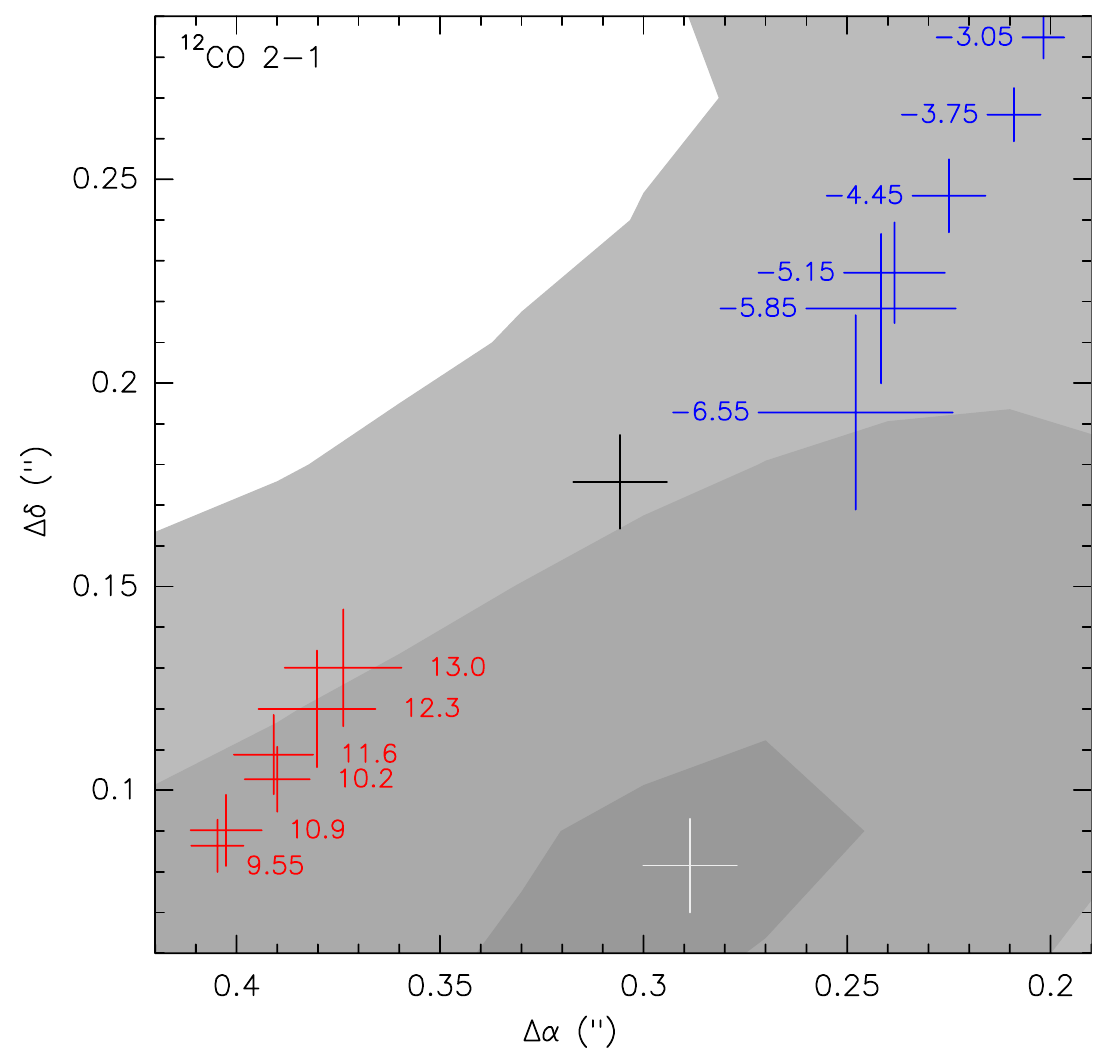}
\caption{Zoomed-in view of the disk center. The gray scale is the dust continuum emission. The color crosses 
are the peak of CO emission at high-velocity channels. The length of the cross indicates the uncertainty of the peak 
position. The numbers next to the crosses indicate the corresponding velocity channel (in km s$^{-1}$). The black cross 
is the averaged center position for the high-velocity channels, and the white cross shows the dust emission peak.
\label{fig7}}
\end{figure}

The highest velocity channels of the CO emission ($V_{lsr} < -3$ \kms for the blueshifted gas and $>10$ \kms for 
redshifted gas) are confined to the disk. We performed Gaussian fits to compute the peak position of these components 
and found that they appear to be located approximately in the same direction (Fig. \ref{fig7}). As the velocity increases, 
the blue and red peak positions move closer to each other in an approximately linear manner, with the averaged position 
indicating the gravitational center. The center of this line indicates that the kinematical center is 14 AU removed from the 
dust peak. 

This offset between the gravitational center and the dust peak indicates that BHB07-11 possibly hosts a multiple system. 
The disk asymmetry could be a barely resolved binary system, since BHB07-11 is a young source \citep[$0.1-0.2$ Myr,][]
{Brooke07} and its disk does not seem to be massive enough to develop gravitational instabilities (Appendix 
\ref{app:model}).

\section{Modeling gravitational instabilities}
\label{app:model}

\begin{figure}[b!]
\centering
\includegraphics[width=\columnwidth,trim={4.2cm 3cm 15cm 0},clip]{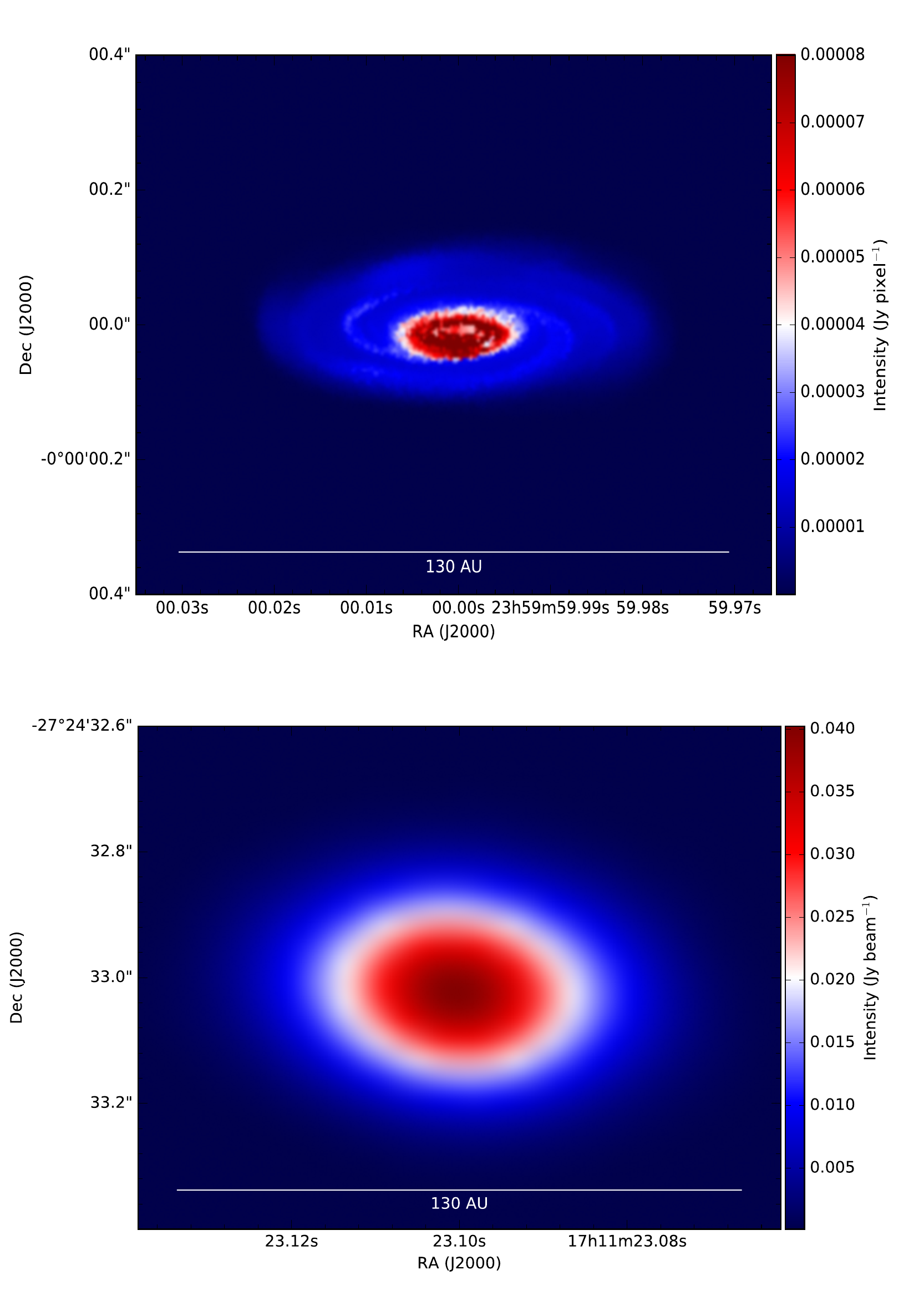}
\caption{{\it Upper panel}: Self-gravitating disk model \citep{EIB15}. The spirals produced by the gravitational instabilities 
are visible at radii larger than 20 AU. {\it Lower panel}: Simulation of this model using the same antenna configuration 
as our observations. The angular resolution of the simulated map is similar to our data ($0\farcs25 \times 0\farcs20$).
\label{fig8}}
\end{figure}

Circumstellar disks with masses comparable to their central protostar can develop gravitational instabilities that are due to their 
self-gravity \citep{Toomre64}. This effect produces spiral density waves that regulate the mass accretion process. In 
order to verify whether the azimuthal asymmetries seen in the disk are due to unresolved spirals, we used a model representing 
an early Class I object with size and mass very similar to the disk in BHB07-11. The model was a self-gravitating disk of 
0.17 \Msun surrounding a solar-type star \citep{EIB15}. The model spanned 130 AU and was placed at the same distance (145 
pc) and inclination (70\degr) as BHB07-11. The peak temperature and density of the model were 330 K and $5.7 \times 
10^{13}$ cm$^{-3}$, respectively, and were confined within a radius of 10 AU in the disk. At larger radii, the temperature 
dropped below 50 K. Through non-local thermodynamic equilibrium radiative transfer processing, using LIME \citep[line 
modeling engine --][]{Brinch10}, we observed that the optical depth varied across the disk structure; it was optically 
thinner in the inner disk. Given the similarities between the physical properties of the model and BHB07-11, we used 
CASA to simulate an observation of this model under the same conditions as our observations; that is, we 
simulated band 6 observations of the model with the same continuum bandwidth as our spectral setup (2.3 GHz) and the 
same antenna configuration. We also added thermal noise using the atmospheric profiles of the ALMA site, including a 
realistic precipitable water vapor level.   

As shown in the lower panel of Fig. \ref{fig8}, the simulations are not able to reproduce the spiral structures because 
they are too small to be resolved in our observations. This implies that a larger and more massive disk than that of 
BHB07-11 is necessary to develop gravitational instabilities that are detectable at 145 pc distance. Even more prominent spirals 
produced by a more massive disk model \citep[0.39 M$_\sun$,][]{IBC11} were not detected in our simulations. Therefore, 
the azimuthal asymmetries must be produced by another mechanism.

\end{appendix}
\end{document}